\documentclass{mem}
\usepackage{natbib}
\usepackage{newtxtext,newtxmath}  
\usepackage{txfonts}
\usepackage{balance}
\usepackage{graphicx}
\usepackage{flushend}
\usepackage{xcolor}
\usepackage[utf8]{inputenc} 
\usepackage[T1]{fontenc} 
\usepackage[version=4]{mhchem}
\usepackage[breaklinks,pdftex]{hyperref}
\idline{75}{282}

\begin{document}

\title{
Computational Astrochemistry
}

\subtitle{Journey towards the molecular universe}

\author{
D. \,Campisi\inst{1} 
\and J. \,Perrero\inst{2,3}
\and N. \,Balucani\inst{4}
}



\institute{
Institute for Theoretical Chemistry, University of Stuttgart, Pfaffenwaldring 55, 70569, Stuttgart, Germany,
\email{campisi@theochem.uni-stuttgart.de, darcampisi@outlook.com}
\and
Departament de Qu\'{i}mica, Universitat Aut\`{o}noma de Barcelona, Bellaterra, 08193, Catalonia, Spain
\and
Dipartimento di Chimica and Nanostructured Interfaces and Surfaces (NIS) Centre, Universit\`{a} degli Studi di Torino, via P. Giuria 7, 10125, Torino, Italy
\and
Dipartimento di Chimica, Biologia e Biotecnologie, Università degli Studi di Perugia, Via Elce di Sotto 8, 06123, Perugia, Italy
\\
}

\authorrunning{Campisi}

\titlerunning{Computational Astrochemistry}

\date{Received: XX-XX-XXXX (Day-Month-Year); Accepted: XX-XX-XXXX (Day-Month-Year)}

\abstract{In astrochemistry, computational methods play a crucial role in addressing fundamental astronomical questions. Interstellar molecules profoundly influence the chemistry and physics of the interstellar medium (ISM), playing pivotal roles in planet formation and the emergence of life. Understanding their chemistry relies on theoretical approaches such as Density Functional Theory (DFT) and post–Hartree–Fock methods, which are essential for exploring pathways to molecular complexity and determining their interstellar abundances. Various theoretical methods investigate the formation of interstellar molecules in both gaseous and solid states. Molecules in interstellar space may originate from bottom-up processes (building up from CO molecules) or top-down processes (polycyclic aromatic hydrocarbon fragmentation). Here, we present a journey of theoretical investigations aimed at studying the reactivity of interstellar molecules in space.

}
\maketitle{}

\section{Introduction}
A wide range of organic molecular species have been observed in different regions of the interstellar medium (ISM) and the solar systems, both within our own galaxy and beyond \citep{McGuire_2022,Cologne_Database}. These molecules serve vital functions in regulating the physical and chemical conditions of the ISM, potentially contributing to the formation of planetary systems and the emergence of life \citep{Tielens:2013}. The launch of the high-resolution James Webb Telescope holds promise for revealing organic molecular species in the interstellar space \citep{JWST:2024}.

\begin{figure}
    \centering
\includegraphics[width=0.5\textwidth]{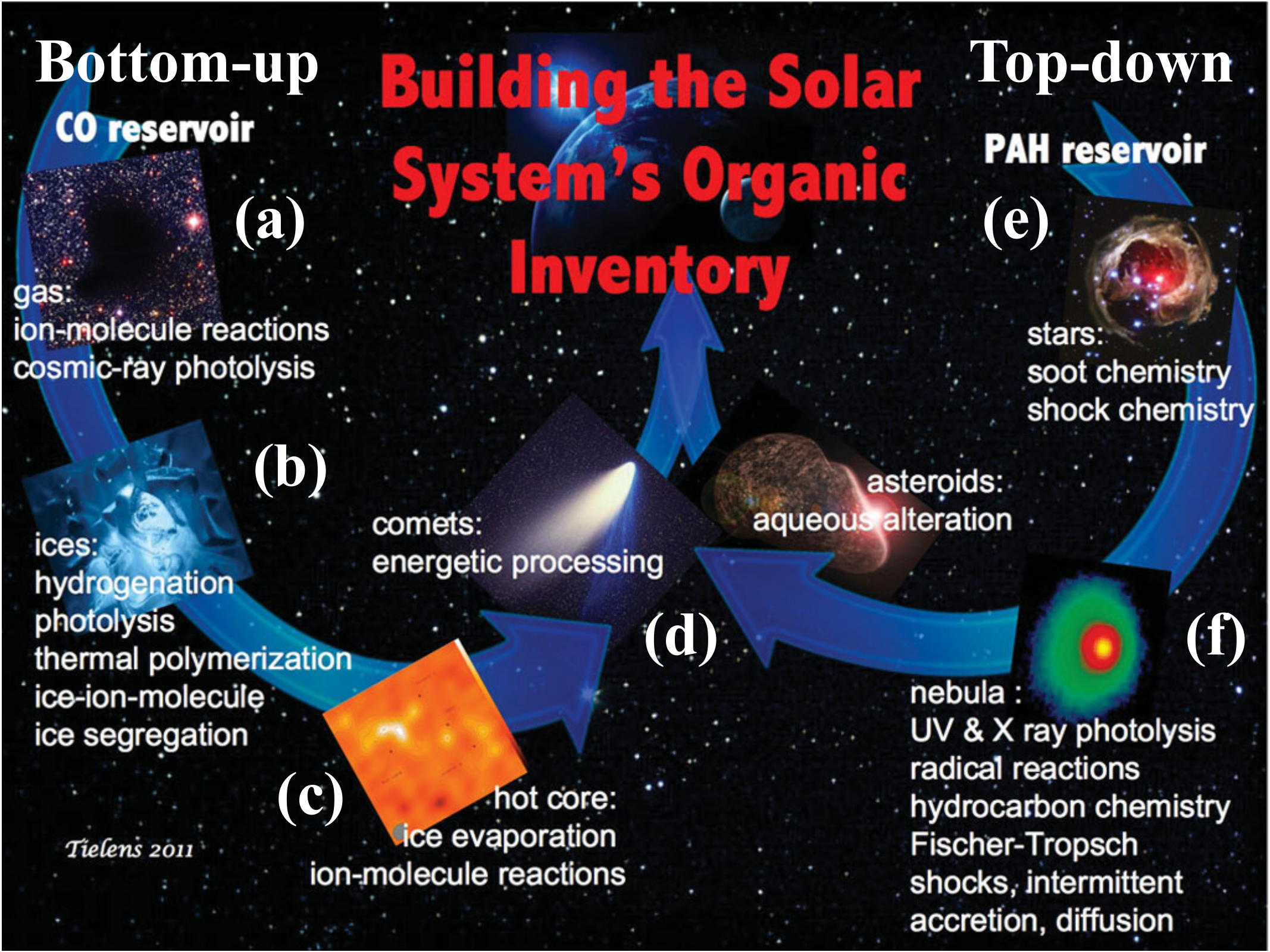}
    \caption{The schematic explanation of the processes of bottom-up and top-down approaches as reported by \cite{Campisi:PhD:2021} and \cite{Tielens:2013}.}
    \label{fig:MolUniverse}
\end{figure}
As the molecules chemically evolve within the ISM, they become integrated into solar system materials like comets, asteroids, and meteorites \citep{Pascale:2000}.
The recent NASA mission OSIRIS-REx successfully returned a sample of the asteroid Bennu to Earth \citep{lauretta_osiris-rex_2017,PARKER202342}. Examination of these solar system materials promises valuable insights into the origin and chemical evolution of molecular species. Isotopic analysis of these materials has proven crucial in tracing the origin of molecules, believed to have originated in the solar nebula \citep{Sandford:2001}. Hence, the molecular species detected in these primitive objects represent a legacy of chemical processes that commenced in the ISM \citep{Tielens:2013}. To elucidate this further, most organic molecules are believed to stem from two prevalent entities in space: carbon monoxide (CO) and Polycyclic Aromatic Hydrocarbons (PAHs) \citep{Tielens:2013}. According to the bottom-up approach, intriguing chemistry occurs on the surfaces of interstellar grains in cold star-forming regions, where CO undergoes hydrogenation, gradually fostering molecular complexity \citep{Tielens:2013}. Conversely, PAHs, constituting approximately 20\% of the carbon in space, are thought to form in the outflows of carbon-rich stars \citep{Tielens:PAH:2008,Mebel:2017}. From a top-down perspective, PAHs fragment via radical and photodissociation processes, producing daughter molecules that act as precursors for biologically relevant molecular structures \citep{Pascale:2006,Tielens:2013}. Eventually, the outcomes of both \textbf{bottom-up} and \textbf{top-down} processes (Fig. \ref{fig:MolUniverse}) become integrated into cometesimals and planetesimals, potentially delivering them to habitable zones and laying the groundwork for life \citep{Tielens:2013}.

Laboratory astrophysics is crucial to understanding the chemical evolution and composition of the cosmos and complements the astronomical approach \citep{ceccarelli2023organic}. 

Due to the challenges posed by certain reactions of interest, conducting studies in terrestrial laboratories becomes exceedingly difficult. These reactions involve transient species, such as radicals or ions, which are challenging to generate in a controlled manner and in sufficient quantities for reactive experiments. Moreover, experimental approaches often struggle to replicate the authentic conditions of the interstellar medium (ISM).
In this context, theoretical calculations play a vital role. They support the interpretation of experimental data and aid in extrapolating results to conditions typical of the ISM. Furthermore, theoretical characterization becomes the sole viable option for investigating reactions that are particularly elusive to study experimentally \citep{ceccarelli2023organic}.

Density Functional Theory (DFT), a prominent quantum chemistry approach \citep{DFT}, efficiently investigates organic molecules in interstellar environments, providing essential spectroscopic constants, binding and activation energies for astronomical data interpretation. Higher-level methods like coupled-cluster (CC) theory \citep{CC} and complete active space perturbation theory (CASPTn) \citep{CASPTn} may be necessary, particularly for neutral-neutral reactions, where minor energy differences in key transition states significantly influence reaction rate coefficients \citep{dawes,balucani2024}. 

Below, we outline common computational strategies for studying interstellar molecule reactivity in the ISM.

\section{Bottom-up approach}
\label{bu}
Early astrochemical models focused on gas-phase reactions to derive the abundances of complex molecules. Initially, it was thought that interstellar dust grains only catalyzed H$_2$ formation, but later, it was proposed that other species could hydrogenate on grain surfaces, increasing molecular complexity \citep{Tielens:2013}. 

The bottom-up formation of interstellar complex organic molecules (iCOMs) involves the production of hydrogenated species on icy grain mantles, which are then desorbed into the gas phase and undergo reactions. Furthermore, radical-radical couplings occur directly on grain surfaces during the warm-up phase of a newly born protostar, aiding in molecular complexity \citep{Tielens:2013}. 

In this section, we analyze the role of computational chemistry in substantiating these scenarios.

\subsection{Reactions on dust grain surface}
Several experimental studies have shown that complex molecules can form on cold ice surfaces (e.g., see \cite{linnartz2016rev,oberg2016photochemistry,kaiser2023}). However, these experiments do not fully replicate the conditions of interstellar ice, such as composition, UV and H atom flux, and grain size. Therefore, a theoretical approach is crucial for understanding these processes at an atomic level and their impact on interstellar ice chemistry \citep{Rimola2018}.

A thorough study of radical-radical recombination reactions on ice surfaces uncovered several challenges \cite{garrod2006,enrique2022APJSS}. First, the coupling of radicals depends on a delicate balance between their diffusion and desorption rates, limiting the temperature range where reactions can occur. For instance, the formation of acetaldehyde from CH$_3$ and HCO is constrained between the temperatures where CH$_3$ can diffuse (9-15K) and when methyl radicals desorb, 30K, \citep{ER2021_rrkm}. Second, while these couplings are often assumed to be barrierless due to coupling of opposite electronic spins, energy barriers can arise as radicals must break surface interactions to diffuse and react. Third, these reactions can have competing channels, like H-abstractions, hindering the formation of important interstellar iCOMs \citep{Tielens:2013}.

An alternative mechanism, the "radical + ice component" scheme, proposes a reaction between a radical (from the gas phase or UV irradiation) and neutral components of ice, like H$_2$O. This approach, discussed by \citet{Rimola2018}, was demonstrated with the formation of formamide (NH$_2$CHO) from CN radical and water on an ice surface. Despite addressing diffusion and competitive reaction issues, this method faces an energy barrier due to the radical-neutral interaction.

The mentioned scheme shows promise, recently tested with CCH, similar in reactivity to CN, on a water ice surface \citep{perrero2022ethanol}. CCH reacts with H$_2$O barrierlessly or with a small barrier, forming vinyl alcohol precursors (H$_2$CCOH and CHCHOH) or C$_2$H$_2$ + OH. Hydrogenation of vinyl alcohol to ethanol has a low activation energy barrier, overcome by tunnelling effects. The formation of C$_2$H$_2$ + OH is hindered, as the water ice surface efficiently dissipates the energy released \citep{pantaleone2020,pantaleone2021}. Nonetheless, this path can effectively generate OH radicals on ice surfaces without direct energy processing.

Quantum chemical evidence supports the "radical + ice component" scheme as an alternative to classical radical-radical recombination mechanisms. This adds to the various proposed reaction mechanisms explaining the presence of iCOMs in the ISM.

\subsection{Gas-phase reactions}
The thorough characterization of gas-phase reactions of relevance in astrochemistry hinges upon three fundamental methodologies: 1) the CRESU (Cinetique de Reaction en Ecoulement Supersonique Uniforme) technique yields low-temperature global rate coefficients \citep{Rowe_1987,Smith:2006}. 2) collision-free experiments determine primary reaction products and their branching ratios \citep{Casavecchia:2015}. Since these techniques are rarely able to reproduce both the low temperature and the low pressure conditions of the interstellar medium, a third approach is necessary based on 3) Quantum theory for chemical reactions aids. This approach is of help in interpreting experimental data and enables extrapolation to interstellar cloud conditions. Additionally, it provides insights for reactive systems inaccessible to experimental investigation.

The quantum characterization of chemical reactions requires electronic structure calculations to derive the potential energy surface (PES) describing the reaction through the stationary point (transition states or bound intermediates) which are formed along the minimum energy path (see, for instance, \cite{dawes} and references therein).
Once the PES of a certain reaction has been determined, kinetics calculations can be performed via transition state theory, variational transition state theory or capture theory \citep{truhlar1984,truhlar1996,clary} depending on the characteristics of the system. In the case of multichannel reactions, a statistical approach can allow to determine the product branching fractions \citep{millerklippenstein}.

As already mentioned above, reactions involving two radical species, like NH + C$_2$H$_5$ (forming mainly CH$_2$NH + CH$_3$) \citep{balucani2018} and O + CH$_2$CH$_2$OH (a key route to interstellar glycolaldehyde) \citep{skouteris2018,vazart2022}, lack experimental characterization and are characterized only theoretically. Radiative-association processes, like those discussed in \citep{nyman1,nymanrev,babb,giani}, are also exclusively theoretically characterized due to challenges in lab experiments: it is difficult to disentangle the stabilization of the addition intermediate by spontaneous emission of photons from collisional stabilization.

Recent astrochemical modeling highlighted a significant case involving CH$_3$O radical in cold or warm regions of the ISM \citep{balucani2024}. The primary formation pathway for CH$_3$O involves the reaction CH$_3$OH + OH, extensively studied experimentally and theoretically \citep{balucani2024}. Understanding this reaction is complex due to its non-Arrhenius behavior, attributed to a pre-reactive complex \citep{shannon2013}. Klippenstein et al. \citep{klippenstein2005,klippenstein2007} showed that for barrierless radical-molecule reactions, the rate-determining step might shift from the formation of a weakly bound van der Waals complex at low temperatures to the passage through a subthreshold saddle point at higher temperatures. The comparison between theoretical calculations and experimental results on several systems suggests a two-transition-states mechanism, necessitating accurate treatments of both inner and outer transition states \citep{klippensteinscience,klippenstein2013}. 

In the CH$_3$OH + OH case, however, the inner transition state lies above the energy of the reactants, creating a real threshold, resulting in slow kinetics at room temperature and above \citep{lin2007,roncero2018,gao2018,stanton2019}. Experimental data from the CRESU technique show a notable rate coefficient increase below 200 K \citep{shannon2013,Antinolo2016}. And surprisingly, the primary product formed with water is the methoxy radical rather than the anticipated hydroxymethyl radical (CH$_2$OH) \citep{shannon2013}. The peculiar behavior is attributed to the pre-reactive complex and efficient tunneling through the outer transition state, facilitating CH$_3$O + H$_2$O formation. At very low temperatures, the long-lived pre-reactive complex allows reaching the product asymptote via barrier tunneling, whereas at high temperatures, the pre-reactive complex's short lifetime makes its impact on the rate coefficient negligible. Theoretical approaches incorporating tunneling and a two-transition-states mechanism replicate CRESU experiment trends \citep{gao2018,stanton2019}. However, experimental and theoretical rate coefficient estimates diverge significantly at the low temperatures relevant to interstellar chemistry \citep{gao2018}.
Partial collisional stabilization of the pre-reactive complex, observed in CRESU experiments, is attributed to conditions incompatible with the highly rarefied ISM environment \citep{gao2018}. 

Intriguingly, comparison between astrochemical model predictions and CH$_3$O observations in varied temperature regions favors theoretical values derived by \cite{gao2018} over experimental data, underscoring the significance of theoretical approaches in simulating interstellar conditions.
\begin{figure}
    \centering
\includegraphics[width=0.5\textwidth]{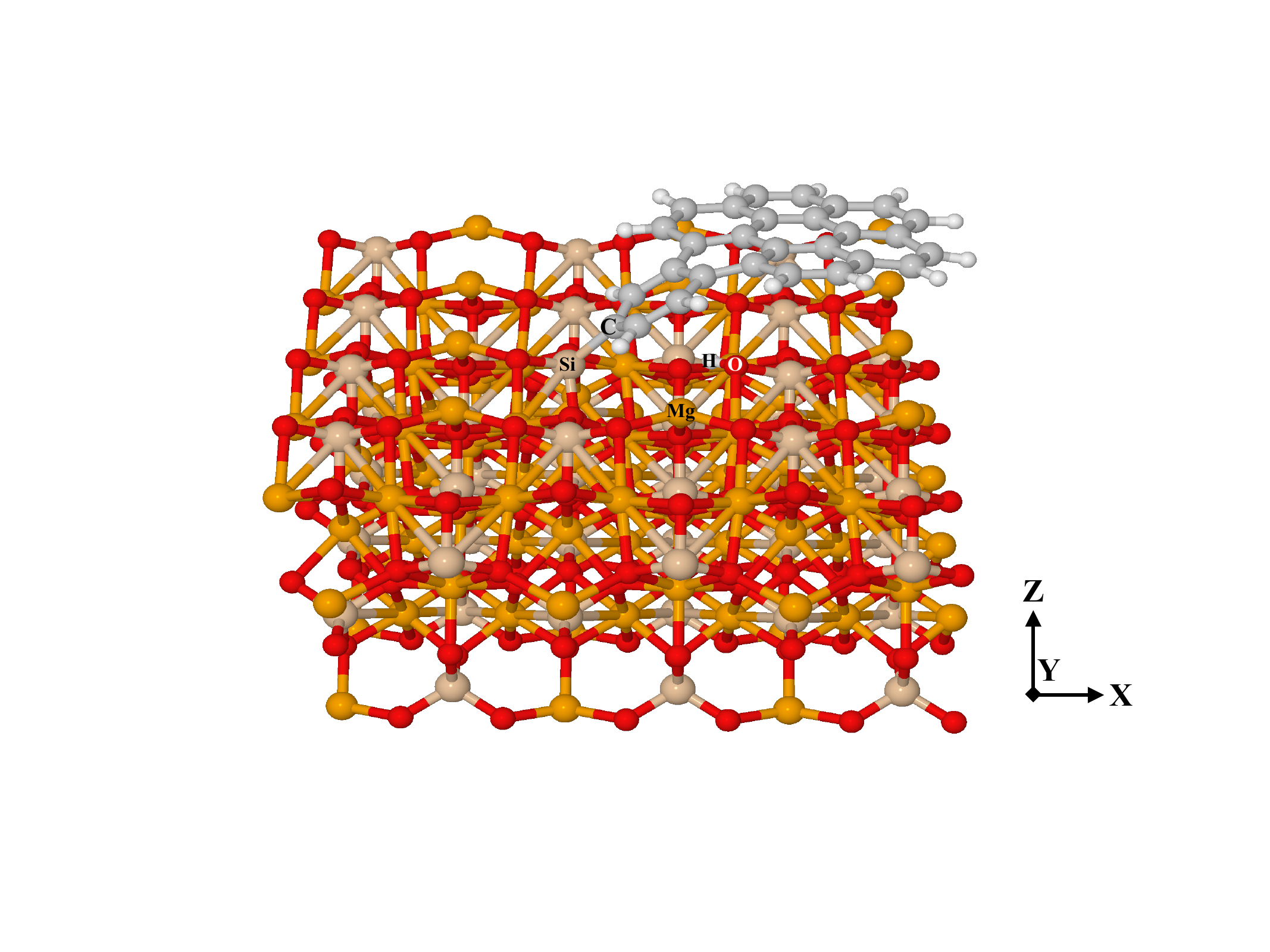}
    \caption{A large PAH, benzocoronene, adsorbed onto a MgO vacancy on the (010) forsterite surface with a binding energy of 5.48 eV \citep{Campisi:2022}. Atomic labels are provided as follows: C (light gray), H (white), O (red), Mg (orange), Si (beige). The surface faces a vacuum region along the z-axis.} 
    \label{fig:PAHs}
\end{figure}
\section{Top-down approach}
\label{td}
\subsection{Interstellar PAHs}
PAHs (Fig. \ref{fig:PAHs}), large carbon molecules abundant in our galaxy and beyond, exist in various forms including individual molecules and solid grains. They are present in solar system objects like carbonaceous chondrites meteorites \citep{Sephton:2002}, believed to preserve ancient organic compounds formed in the solar nebula. These materials offer insight into early solar system chemistry. PAHs smaller than 6-membered rings are detected in meteorites, suggesting their role as carbon reservoirs and potential association with prebiotic molecules \citep{Tielens:PAH:2008,Tielens:2013}.

After being formed in the surroundings of carbon-rich stars, PAHs are injected into the ISM, where various chemical and photochemical processes can occur \citep{PAHsoot:1989,Cherchneff:1992}. Thus, Density Functional Theory (DFT) has been extensively used to study both the spectroscopy \citep{IR_PAHs} and reactivity (\cite{Jensen:2019,Campisi:2020} and many others) of PAHs in different environments, such as when adsorbed onto grain surfaces or present as gas-phase molecules. The main distinction lies in the treatment of the theoretical model, with periodic DFT \citep{Philip:2014} taking into account periodic potentials to simulate solid boundary conditions.

Periodic DFT was used to study PAH adsorption on forsterite surfaces, common in interstellar grains and planetary systems \citep{campisi:2021,Campisi:2022}. Forsterite is found in carbonaceous chondrite asteroids and meteorites where PAHs are detected \citep{Sephton:2002}. Large PAHs strongly adsorb onto forsterite due to non-covalent aromatic ring interactions (Fig. \ref{fig:PAHs}), with binding energies ranging from 1 eV for naphthalene to about 5 eV for benzo-coronene \citep{Campisi:2022}. Defects like MgO vacancies enhance reactivity by providing cavities for PAHs to intercalate and dissociate their C-H bonds, potentially explaining the absence of large PAHs in solar system materials.

PAHs play a crucial role in the interstellar medium (ISM) by hosting radicals and undergoing superhydrogenation, a process essential for catalyzing molecular hydrogen formation, the most abundant molecule in space \citep{Jensen:2019,Campisi:2020,Rauls_2008}. Non-periodic DFT studies have elucidated the mechanism of PAH superhydrogenation, indicating that odd hydrogenation attachments have barriers approximately 0.20 eV due to involvement of radical neutral species, while even hydrogenation proceeds barrierlessly through radical-radical recombination \citep{Jensen:2019,Campisi:2020}.
PAH reactivity is shape-dependent, with linear PAHs—where aromatic rings are connected in a straight line—being more reactive than larger, more condensed PAHs. DFT is essential for understanding experimental mass spectra measurements, revealing magic numbers (numbers of hydrogenated species) corresponding to barriers higher than 0.20 eV \citep{Jensen:2019,Campisi:2020}. Superhydrogenation not only aids in molecular hydrogen formation but also weakens carbon-carbon bonds in PAHs, converting aromatic bonds into aliphatic ones. Gas-phase studies on cationic pyrene \citep{Tang:2022} indicate that ethyne extraction via fragmentation is possible, but barriers exceed 1.50 eV. Comparing gas-phase studies to reactions on solid surfaces is challenging due to energy dissipation by the grain during exothermic reactions, resulting in a lack of internal energy to overcome the significant barriers.

N($^2$D) and O($^3$P) are highly reactive toward hydrocarbon molecular species \citep{Balucani:2023,Rosi:2020,Cavallotti:2020}, crucial for life-building blocks. Current research focuses on their reactivity with PAHs \citep{Aponte:2017}. Computational studies have revealed N($^2$D) interactions with small hydrocarbons, leading to N-heterocycle rings and HCN formation, important in Strecker amino acid reactions \citep{Balucani:2023,Aponte:2017}. O($^3$P) shows promise for PAH fragmentation \citep{Cavallotti:2020}, with both experimental and computational studies identifying pathway channels in triplet and singlet states, resulting in CO and small ketene formation \citep{Cavallotti:2020,Dulieu:2019}. Further research is needed to explore how PAH size influences O atom reactivity.

Computational studies are pivotal in assessing the photodissociation rate of PAHs by examining their excitation in the UV-visible range (\citep{Joblin_2020,Marciniak:2021} and many others). PAHs are susceptible to photo-destruction, especially in environments like active galactic nuclei or near massive stars, where they can undergo H-loss, fragmentation, and ionization \citep{Zhen_2015}. Additionally, collision heating from interactions with cosmic rays, heavy and fast atoms, ions, and shock waves can further contribute to PAH fragmentation by interacting with energetic particles (\citep{Allamandola:1989} and many others).

\section{Conclusions}
Theoretical methods play a crucial role in studying the chemical evolution reactivity prediction of interstellar molecules.
From a bottom-up perspective, complex organic molecule synthesis (iCOMs) results from the collaboration of on-grain and gas-phase reactions. On-grain reactions, like radical-radical couplings, face minimal activation barriers but are constrained by the delicate balance of reactant diffusion and surface desorption. Gas-phase reactions present challenges in determining accurate rate coefficients and simulating realistic interstellar conditions.

Understanding PAH fragmentation is crucial for comprehending the formation of iCOMs, given their high carbon content. Key factors include radical reactions with hydrogen, oxygen, and nitrogen, as well as radiative and ionization processes. Future research should prioritize studying PAH fragmentation induced by oxygen and nitrogen attachment and determining the abundance of resulting products in space.

Moreover, future research should evaluate the delivery of iCOMs formed from both bottom-up and top-down chemistry to understand their presence in solar system objects and their connection with the building blocks of life.

\begin{acknowledgements}
D.C. acknowledges the Alexander von Humboldt Foundation for financial support and Johannes Kästner for fruitful discussions and for computational support.  
J.P. acknowledges financial support from European Union’s Horizon 2020 research and innovation programme from the European Research Council (ERC) for the projects ``Quantum Chemistry on Interstellar Grains” (QUANTUMGRAIN).  
N.B. thanks the Italian MUR (PRIN 2020, “Astrochemistry beyond the second period elements”, Prot. 2020AFB3FX) for support.

\end{acknowledgements}

\bibliographystyle{aa}
\bibliography{Article}

\end{document}